\documentclass[letterpaper,twocolumn,10pt]{article}
\usepackage[twoside=true, head=13pt,
     paperwidth=8.5in, paperheight=11in,
     includeheadfoot=false, columnsep=2pc,
     top=1in, bottom=1in, inner=0.75in, outer=0.75in,
     marginparwidth=2pc,heightrounded]{geometry}

\usepackage{graphicx}
\usepackage[tt=false, type1=true]{libertine}
\usepackage{mathptmx}
\usepackage{amssymb}
\usepackage[varqu]{zi4}
\usepackage[T1]{fontenc}
\usepackage{enumitem}
\usepackage[font={footnotesize},labelfont={footnotesize,bf},textfont={footnotesize,it}]{caption}
\setlist[itemize]{leftmargin=*}

\usepackage{titlesec}
\usepackage{listings}
\usepackage{xcolor}
\usepackage{float}
\usepackage{enumitem}
\usepackage{mathtools}
\usepackage{dblfloatfix}

\usepackage{booktabs}
\usepackage{pifont}
\newcommand{\cmark}{\ding{51}}  
\newcommand{\xmark}{\ding{55}}  

\usepackage[mode=tex]{standalone}
\usepackage{tikz}
\usepackage{tikz-qtree}
\usetikzlibrary{fit,positioning,shapes,arrows.meta}

\usepackage[colorlinks=true, allcolors=black]{hyperref}

\titlespacing*{\section}{0pt}{1.6ex}{1.4ex}
\titlespacing*{\subsection}{0pt}{1.4ex}{1.4ex}

\titleformat{\paragraph}[runin]{\normalfont\normalsize\bfseries}{}{}{}[.]
\titlespacing*{\paragraph}{0pt}{0.6ex plus .2ex minus .1ex}{0.4em}

\def\Plus{\texttt{+}}

\PassOptionsToPackage{destlabel}{hyperref}

\setlist[itemize]{leftmargin=*}

\definecolor{keywordgreen}{rgb}{0.549, 0.698, 0.435}
\definecolor{keywordpurple}{rgb}{0.4,0,0.6}
\definecolor{blockblue}{rgb}{0,0.4,0.8}

\lstdefinelanguage{bigrapher}{
  keywords={react, begin, end, init, rules},
  morekeywords=[2]{Room, Person, CtrlPanel, fix_secure, leave_room},
  sensitive=true,
  morecomment=[l]{//},
  morestring=[b]``,
}

\lstdefinestyle{bigrapherStyle}{
  backgroundcolor=\color{white},
  breaklines=true,
  frame=lines,
  numberstyle=\tiny\color{gray},
  basicstyle=\footnotesize\ttfamily,
  keywordstyle=\color{black}\bfseries,
  keywordstyle=[2]\color{blockblue},
  commentstyle=\itshape\color{purple!40!black},
  breaklines=true,
  breakatwhitespace=true,
  showstringspaces=false,
  tabsize=2,
  columns=fullflexible,
  keepspaces=true,
  linewidth=\columnwidth
}

\colorlet{punct}{red!60!black}
\definecolor{background}{HTML}{EEEEEE}
\definecolor{delim}{RGB}{20,105,176}
\colorlet{numb}{magenta!60!black}

\lstdefinelanguage{json}{
    basicstyle=\normalfont\ttfamily,
    stepnumber=1,
    numbersep=8pt,
    showstringspaces=false,
    breaklines=true,
    literate=
     *{0}{{{\color{numb}0}}}{1}
      {1}{{{\color{numb}1}}}{1}
      {2}{{{\color{numb}2}}}{1}
      {3}{{{\color{numb}3}}}{1}
      {4}{{{\color{numb}4}}}{1}
      {5}{{{\color{numb}5}}}{1}
      {6}{{{\color{numb}6}}}{1}
      {7}{{{\color{numb}7}}}{1}
      {8}{{{\color{numb}8}}}{1}
      {9}{{{\color{numb}9}}}{1}
      {:}{{{\color{punct}{:}}}}{1}
      {,}{{{\color{punct}{,}}}}{1}
      {\{}{{{\color{delim}{\{}}}}{1}
      {\}}{{{\color{delim}{\}}}}}{1}
      {[}{{{\color{delim}{[}}}}{1}
      {]}{{{\color{delim}{]}}}}{1},
}

\lstdefinestyle{jsonStyle}{
  backgroundcolor=\color{white},
  breaklines=true,
  frame=lines,
  breaklines=true,
  breakatwhitespace=true,
  showstringspaces=false,
  tabsize=2,
  columns=fullflexible,
  keepspaces=true,
  linewidth=\columnwidth
}

\floatstyle{plain}
\newfloat{program}{htbp}{lop}
\floatname{program}{Figure}

\DeclareUnicodeCharacter{2212}{-}

\PassOptionsToPackage{destlabel}{hyperref}

\begin{document}

\title{\textbf{An Architecture for Spatial Networking}}
\author{
  \begin{tabular}{ccc}
    \begin{tabular}{c}
      {\Large Josh Millar}\thanks{Both Josh Millar and Ryan Gibb contributed equally to this work.} \\[2pt]
      \normalsize Imperial College London
    \end{tabular}
    & \hspace{2em}
    \begin{tabular}{c}
      {\Large Ryan Gibb}\protect\footnotemark[1] \\[2pt]
      \normalsize University of Cambridge
    \end{tabular}
    & \hspace{2em}
    \begin{tabular}{c}
      {\Large Roy Ang} \\[2pt]
      \normalsize University of Cambridge
    \end{tabular}
    \\[16pt]
    \multicolumn{3}{c}{%
      \begin{tabular}{cc}
        \begin{tabular}{c}
          {\Large Anil Madhavapeddy} \\[2pt]
          \normalsize University of Cambridge
        \end{tabular}
        & \hspace{2em}
        \begin{tabular}{c}
          {\Large Hamed Haddadi} \\[2pt]
          \normalsize Imperial College London
        \end{tabular}
      \end{tabular}
    }
  \end{tabular}
}
\date{}

\maketitle

\begin{abstract}
  Physical spaces are increasingly dense with networked devices, promising seamless coordination and ambient intelligence.
  Yet today, cloud-first architectures force all communication through wide-area networks regardless of physical proximity.
  We lack an abstraction for spatial networking: using physical spaces to create boundaries for private, robust, and low-latency communication.
  We introduce \textit{Bifr\"ost}, a programming model that realizes spatial networking using bigraphs to express both containment and connectivity, enabling policies to be scoped by physical boundaries, devices to be named by location, the instantiation of spatial services, and the composition of spaces while maintaining local autonomy.
  Bifr\"ost enables a new class of spatially-aware applications, where co-located devices communicate directly, physical barriers require explicit gateways, and local control bridges to global coordination.
\end{abstract}

\section{The Spatial Disconnect}
\label{sec:intro}

Physical containment creates a natural network hierarchy, yet we do not currently take advantage of this.
Even local interactions between devices often require traversal over a wide-area network (WAN), with consequences for privacy, robustness, and latency.
Instead, devices in the same room should communicate \textit{directly}, while physical barriers should require explicit networking gateways.
We call this \textit{spatial networking}: instead of overlaying virtual addresses over physical network connections, we use physical spaces to constrain virtual network addresses.
This lets users point at two devices and address them by their physical relationship; devices are named by their location, policies are scoped by physical boundaries, and spaces naturally compose while maintaining local autonomy.

Consider a multi-site meeting in a secured conference room.
As participants arrive, the digital lock recognizes their presence and grants access, while temporary access for an external consultant remains valid only for the meeting duration.
As the room adapts to each arrival and transcription begins at quorum, colleagues gathering at a remote office trigger a secure tunnel between these two spaces, sharing displays as if both rooms were one.
When an unexpected visitor accidentally enters, everything blanks instantly within the ``uncanny valley of human perception''~\cite{osmose}, resuming only after they leave.
When the meeting ends, ephemeral components dissolve: the consultant's access expires, the cross-site tunnel closes, and the transcription service vanishes.

Today's systems cannot achieve this vision.
Cloud platforms route even room-level interactions through WANs: data trails expose movement patterns, Internet outages break robustness---like doors that won't unlock~\cite{nyt2021facebook}---and spatial reactions suffer hundred-millisecond delays.
Local hubs improve privacy, robustness, and latency but ignore physical boundaries as natural security perimeters.
Two devices in different rooms share the same flat network; a compromise in one room threatens all, requiring complex policy to recreate the isolation that walls already provide.
When remote colleagues join meetings, it's all or nothing: Virtual Private Network (VPN) access to the entire building or manual port forwards per device, because rooms can't selectively share their resources.
Each building's hub is an isolated island; your preferences don't travel with you.
Hubs can't delegate to rooms: limiting a transcription service to a room's microphone requires manual access control because the network doesn't respect room boundaries.
Without physical boundaries, we lose natural isolation and need brittle manual configuration to recreate what the space already provides.

We propose \textit{Bifr\"ost}, a spatial networking system that realizes this vision through bigraphs, a two-decade-old formalism that has remained largely theoretical~\cite{milner2008bigraphs}.
Bigraphs simultaneously express containment (where things are) and connectivity (how they interact).
They were presented primarily as a mathematical framework with heavy focus on formal semantics, offering little compelling application beyond encoding other process calculi, and existing tools targeted formal verification.
Ensuing research has largely followed the same trajectory, refining the formalism but leaving it unchanged in ways that might make it practical for systems design.
This persistent abstraction-first focus may help explain why bigraphs, despite their conceptual fit, never found traction as a substrate for runtime systems.

With Bifr\"ost, networking follows spatial boundaries: room-level for immediate interactions, building-level for coordination, WAN only when explicitly bridged.
This spatial approach enables us to write short policies as \textit{reaction rules}---pattern-match-rewrite rules that fire when spatial conditions hold---leveraging physical isolation for security while selectively opening channels for collaboration.
Because these rules are tied to spatial predicates like occupancy or proximity, services can be instantiated just-in-time when conditions hold and automatically dissolved when they no longer apply, with full lifecycle management of network resources---Domain Name System (DNS) names, Transport Layer Security (TLS) certificates, and WireGuard tunnels.

Spatial networking introduces the following:
\begin{itemize}
  \item \textbf{Physical boundaries as network boundaries}: Physical containment naturally creates network isolation; no complex firewall rules or access control lists needed.
  \item \textbf{Automatic spatial naming}: We use a device's location in the bigraph to assign it a name in the Spatial Name System~\cite{sns}, which extends the DNS with hierarchical location-based names, with split-horizon resolution, giving devices stable identity through hardware changes while enabling both local and global reachability (\S\ref{sec:spatial-names}).
  \item \textbf{Spatial service instantiation}: Services exist only when and where needed, spawned by spatial predicates and dissolved when conditions change (\S\ref{sec:spatial-services}).
  \item \textbf{Spatial composition}: Rooms maintain autonomy while opening explicit apertures for collaboration, bridging local and global as needed.
\end{itemize}

The remainder of this paper analyzes the limitations of cloud and local architectures (\S\ref{sec:cloud}), shows how bigraphs capture physical spaces (\S\ref{sec:spatial}), develops our programming model with reaction rules and spatial services (\S\ref{sec:programming}), demonstrates spatial networking through concrete scenarios (\S\ref{sec:spatial-networks}), details our runtime implementation (\S\ref{sec:impl}), evaluates system performance (\S\ref{sec:eval}), positions our work within related research (\S\ref{sec:related}), and concludes with future directions (\S\ref{sec:conclusions}).

\section{The Here and There}\label{sec:cloud}

Cloud platforms achieve \textit{global coordination} through architectural centralization: devices establish outbound connections to cloud servers that broker all interactions.
This model excels at certain tasks---remote management, data aggregation, cross-site synchronization---where global visibility matters more than local responsiveness.
But consider what happens to our meeting scenario from \S\ref{sec:intro} in the cloud.
Room-level interactions are forced through wide-area networks: when an unexpected visitor enters, presence detection must travel to distant servers and back, adding hundreds of milliseconds to what should be instant.
The meeting transcript leaves the room by design, processed and stored in data centers you don't control.
When the Internet fails, the entire space becomes inoperable---doors won't unlock, displays freeze, meetings halt---despite all hardware functioning perfectly.
The cloud's strength is global coordination, but it achieves this by erasing locality.

\begin{figure*}
  \centering
  \includegraphics[width=.98\textwidth]{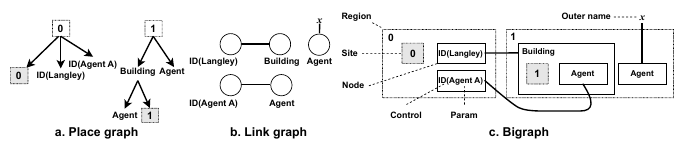}
  \caption{A \textit{bigraph}, constructed from its underlying \textit{place} and \textit{link} graphs.}\label{fig:fig2}
\end{figure*}

Local hubs improve on the cloud's \textit{privacy} and \textit{robustness}: data stays within the network perimeter~\cite{hass, openhab}, devices work during Internet outages, and \textit{latency} drops from WAN round-trips to local-area network (LAN) milliseconds.
But building-level is still too coarse---the flat network ignores room boundaries, so securing a private meeting requires manual virtual LANs (VLAN) and firewall rules to recreate the isolation that walls already provide.
These systems are fundamentally autarkic, with no mechanism for controlled delegation to the room-level, where most interactions actually occur.
To achieve the meeting scenario from \S\ref{sec:intro}, for example, administrators must manually configure VPN tunnels, firewall rules, and port forwards for each interaction---brittle plumbing that cannot express the spatial, temporal, or hierarchical nature of the requirement.
What we need is \textit{spatial composition}: room-level granularity for immediate interactions, building-level for coordination, and WAN bridging only when necessary.

\begin{table}[b]
\centering
\setlength{\tabcolsep}{6pt}
\renewcommand{\arraystretch}{1.2}
\resizebox{6cm}{!}{
\begin{tabular}{@{}lccc@{}}
\toprule
\textit{Capability} &
\textbf{Cloud} &
\textbf{Hub} &
\textbf{Bifr\"ost} \\
\midrule
Global coordination  & \cmark & \xmark & \cmark \\
Privacy              & \xmark & \cmark & \cmark \\
Robustness           & \xmark & \cmark & \cmark \\
Low-latency          & \xmark & \cmark & \cmark \\
Spatial composition  & \xmark & \xmark & \cmark \\
\bottomrule
\end{tabular}}
\caption{Architectural capabilities.
  Only Bifr\"ost provides spatial composition alongside both global coordination and local autonomy.}
\label{tab:comparison-bifrost}
\end{table}

Both cloud platforms and local hubs fail to realize this because they cannot express spatial relationships as first-class constructs.
We need virtual addressing based on physical location: when two devices are in the same room, they can communicate directly; when they're in different spaces, an explicit aperture must be opened.
Physical containment becomes the network boundary, eliminating complex firewall rules and VLANs because isolation emerges naturally from spatial structure.
This spatial composition allows autonomous rooms to nest within floors and buildings, each maintaining boundaries while selectively bridging when needed, with policies that compose hierarchically through the containment structure.
The timing is right: smartphones now capture detailed models of interiors (\S\ref{sec:mapping:roomplan}), and a growing class of devices are inherently spatial---displays, locks, thermostats, and sensors that never move and exist to serve the space they occupy (\S\ref{sec:spatial-names}).
These devices should be named by their spatial identity: the display in room 101 \textit{is} ``display.room-101'', stable and meaningful regardless of hardware replacements.
This enables spaces to instantiate ephemeral services under similarly stable names: ``transcribe.room-101'' exists exactly when room 101 hosts a meeting.
Bifr\"ost realizes spatial networking through bigraphs, which express both containment (the \textit{here}) and connectivity (the \textit{there}) in a single, programmable model.

\section{Defining Physical Spaces \\ with Bigraphs}\label{sec:spatial}

To realize spatial networking, Bifr\"ost treats \textit{space} as a first-class programming dimension, rather than gluing together ad hoc automations across device IDs,
Our model exposes a structured object---the \textit{bigraph}---that simultaneously captures \textit{where} things are (containment, adjacency, boundaries) and \textit{how} they can interact (links, communication, authority).
Policies, behaviors, and services are expressed as rewrites that react to spatial context.

\subsection{Bigraphs as a Spatial Substrate}

A bigraph~\cite{milner2008bigraphs} consists of two orthogonal structures over the same set of typed entities (nodes): \textit{place} and \textit{link}.
Figure~\ref{fig:fig2} shows a worked example.

The place graph $P$ is a rooted forest encoding nested spatial regions.
Roots model distinct regions (e.g., \texttt{Building~A}, \texttt{Campus~B}); children encode containment (e.g., building $\rightarrow$ floor).
Logical regions (e.g., ``hotdesk area'') are modeled as places too, making containment an explicit, programmable property rather than an implicit assumption baked into node names or network subnets.

The link graph $L$ is a hypergraph connecting \textit{ports} on nodes with (possibly open) hyperedges.
Links capture non-spatial relationships: communication channels, social/organizational ties, or memberships (e.g., the \texttt{AV\_LAN} link joining a display and microphone).
\textit{Open links} terminate at \textit{outer names} that serve as the boundary interface between subgraphs; they are the ``seam'' when composing subgraphs across which delegated hosts synchronize.

Each node is labeled with a \textit{control} (a type) and a bounded number of ports defined by the control.
Controls act as a schema: \texttt{User}, \texttt{Display}, \texttt{Network}, etc.
Ports make link intent explicit (e.g., \texttt{Display.video\_in}); a link between ports conveys a capability to interact.

Subgraphs advertise their interface through roots (place boundaries) and open links (outer names).
Composition is by \textit{nesting}  (plugging a subgraph into a place) or by \textit{parallel} composition (placing disjoint subgraphs side-by-side).
This means a policy authored once for a subgraph is portable to any other subgraph with the same interface.

\subsection{Mapping Location \& Space}\label{sec:mapping}

A common practical concern with spatial programming is the effort required to assemble robust models of the environment, especially for large offices or campuses.
In practice, however, this is fairly manageable: using off-the-shelf data, we can assemble accurate, useful graphs with modest effort.
To make this concrete, we ship ingestion pipelines for three widely available sources to bootstrap initial deployments; (1) an OpenStreetMap (OSM) importer that assembles a world-to-campus hierarchy; (2) a mobile RoomPlan-based pipeline that turns commodity iPhone/iPad 3D scans into typed indoor subgraphs, and (3) an OwnTracks-based importer that maps live user locations into the place graph.

\paragraph{Bootstrapping from OSM}
Our OSM importer directly maps primitives (nodes/ways/relations) and tags into controls and boundaries.
Operationally, detailed administrative areas, building footprints, and (when present) indoor tagging form the place hierarchy (see Figure~\ref{fig:office}); salient tags (e.g., \texttt{building=office}, \texttt{level=1}, \texttt{name}) become typed properties; entries/exits are exposed as outer names to compose with nested indoor subgraphs.
Our importer supports bounded fetches (by place or bounding box) and incremental updates; merges are idempotent, so long-running deployments track upstream edits without manual rebasing.
This yields a “world $\rightarrow$ campus $\rightarrow$ building” skeleton into which indoor models can be nested.

\begin{figure}[t]
  \centering
  \includegraphics[width=\linewidth]{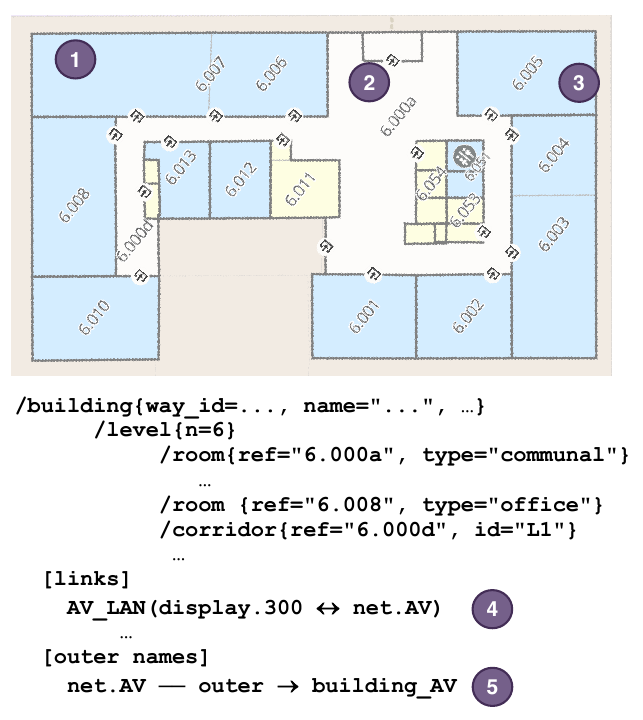}
  \caption{Compiling OpenStreetMap (OSM) and ancillary data into a bigraph.
    (1) OSM building relation/way becomes the root of the place graph, with tags (e.g., \texttt{way\_id}, \texttt{name}) stored as properties; levels (\(n=6\)) form the next layer.
    (2) A specific room (e.g., \texttt{room\{ref="6.008", type="office"\}}) is compiled as a node with its OSM tags as properties.
    (3) Each node is assigned a stable name for addressing (e.g., \texttt{6.008}, \texttt{corridor L1}).
    (4) The link graph captures connectivity, e.g., \texttt{AV\_LAN(display.300)} $\leftrightarrow$ \texttt{net.AV}.
    (5) We export select links as outer names to form controlled apertures across boundaries, e.g., \texttt{net.AV} $\to$ \texttt{building\_AV}.}
  \label{fig:osm-office}
\end{figure}

\paragraph{Mapping Interiors with RoomPlan}\label{sec:mapping:roomplan}

To rapidly capture interiors, we ship a RoomPlan~\cite{roomplan} parser that ingests the JSON produced by a commodity iPhone/iPad scan.
The mapped layout builds the place graph; poses and extents are kept as typed geometry (units, frames), and recognized nodes map to known controls.
This includes device nodes: during capture, a user can indicate the location and type of equipment (e.g., displays, switches), and the parser inserts them into the graph.
The parser also derives links for adjacency and capabilities (e.g., a \texttt{Display.video\_in} port linked to an \texttt{AV\_LAN} hyperedge), de-duplicates near-coincident edges, and normalizes coordinates.
Each indoor subgraph is anchored under its OSM building or campus/site parent using geodetic hints.
Repeat captures are supported; stable GUIDs let merges update nodes in place.

\begin{figure}[t]
  \centering
  \includegraphics[width=6.8cm]{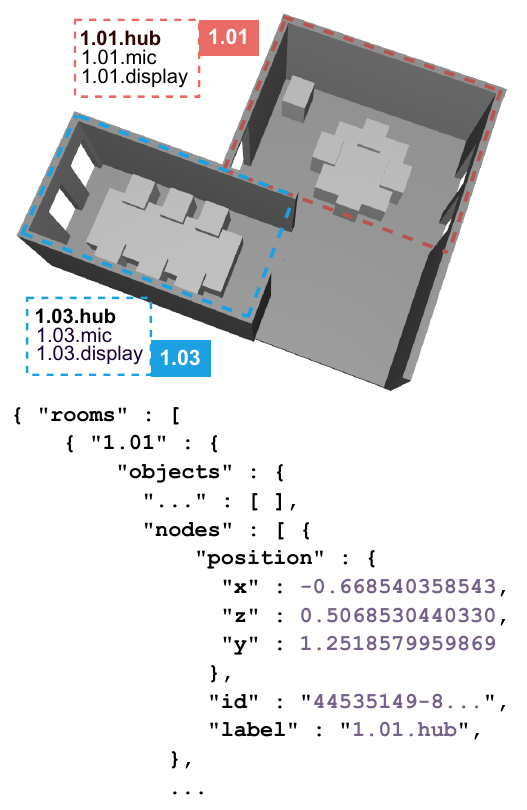}
  \caption{A small 3D office map, generated on iPhone using Apple's RoomPlan API, with its corresponding JSON parsing.
  }\label{fig:office}
\end{figure}

\paragraph{Live Location Tracking}

To reflect occupancy in real time, we must ingest user/device positions into the place hierarchy.
Outdoor positioning is relatively straightforward; indoor positioning varies by building and infrastructure readiness.

Outdoors, global navigation satellite systems (GNSS) suffice.
We implement an OwnTracks\footnote{\url{https://owntracks.org/}} MQTT pipeline to map a user's GPS fixes into the place hierarchy.
We use reverse-geocoding\footnote{\url{https://nominatim.org/}} to resolve latitude/longitude to the nearest OSM element, then map the element to its corresponding place node and reparent the user under it.
Where OSM data is sparse, administrators can use custom geofences.

Within a building, however, GNSS degrades so deployments rely on indoor positioning systems (IPS).
Practical options include widely used radio-based methods (e.g., Wi-Fi/BLE fingerprinting), infrastructure-assisted systems (e.g., badge readers or active beacons), and commercial IPS that use other sensory information to locate objects within a building~\cite{tariqNonGPSPositioningSystems2017}.
Early systems like Active BAT demonstrate fine-grained room/zone localization with active tags~\cite{harterAnatomyContextAwareApplication2002}.
Which technique is used is site-dependent; our model handles heterogeneous inputs uniformly.
The key point is portability: any source yielding {position, accuracy, timestamp} drops into the same pipeline.

By design, our common schema makes it straightforward to ingest structure, geometry, and live location from the sources above or others (e.g., Wi-Fi/BLE fingerprinting, LLM-parsed floorplans, or manual annotations).
Because rules operate on controls and properties---not node-specific IDs---administrators can mix sources and update or replace them over time without changing policy.

\section{Networking with Bigraphs}\label{sec:programming}

Having established how bigraphs represent physical spaces, we now program behaviors over these spatial structures through \textit{reaction rules} (\S\ref{sec:reactions-over-space}): pattern-match-rewrite rules that fire when spatial conditions hold, transforming static graphs into dynamic environments.
We extend bigraphs with typed properties for state and reaction rules with side-effects for practical use (\S\ref{sec:properties-effects}).
Using reaction rules we can create ephemeral services with spatial lifecycles (\S\ref{sec:spatial-services}), and using the place graph we can create hierarchical DNS naming (\S\ref{sec:spatial-names}).
Together, these enable declarative spatial policies that are concise, portable, and composable---a rule written for one room works in any room with the same interface.

\subsection{Reactions over Space}\label{sec:reactions-over-space}

\textit{Reaction rules} are our operational model; they match a spatial pattern and rewrite it, possibly with \textit{side-effects} (\S\ref{sec:properties-effects}).
Patterns can assert, for example, that certain nodes are co-located, that no unauthorized node appears in a subplace, or that specific connectivity holds.

Guards are predicates over these bound variables and node properties, such as time-of-day, user role, or service state.
Guards range from arithmetic checks to lookups encapsulated as property queries (e.g., checking authorization flags).
Importantly, guards are always evaluated within a spatial scope: a rule attached to a room only sees that room's subgraph, not the entire building.

A successful rule application produces an updated subgraph; it may create/remove/move nodes in the place graph, add/remove links, set properties, etc.
For example, the rule in Listing \ref{listing:1} spawns an ephemeral service when a quorum is reached; and on exit, a complementary rule dissolves it and retracts exposure.
This rule is scoped to a room, never hardcoding hostnames or device IDs.
The same rule can be transported by composition to any room that exposes the same open links and controls.

\label{listing:1}
\begin{lstlisting}[language=bigrapher, style=bigrapherStyle, label={lst:quorum}, captionpos=b, caption={Rule to spawn a spatially-scoped transcription service.}]
react start_transcription =
  ( MeetingRoom{label="1.01"}.
      ( Persons{n>=q, authorized=true} || Mic || ... )
        || Net{kind="av", outer=`intra_av`} )
-->
  ( MeetingRoom{...}.
      ( Persons{...} || Mic
        || TranscriptionService{state="active"}
        || ... )
        || Net{kind="av", outer=`intra_av`} ),
  effect spawn_container(name=`transcribe.room-101`),
  effect expose_dns(name=`transcribe.room-101`);
\end{lstlisting}

\subsection{Properties and Effects}
\label{sec:properties-effects}

We extend the classic bigraph formalism in two ways.

\paragraph{Properties} Each node exposes a typed dictionary of properties (e.g., \texttt{authorized=true}).
Properties let policies depend not only on dynamic state, such as occupancy or sensor readings, but also on structure.
They can be initialized from simple YAML templates.
In addition, nodes may carry \textit{geodetic} metadata---\texttt{geom} (Point/Polygon/MultiPolygon), \texttt{crs} (e.g., \texttt{EPSG:4326} or a local floor CRS), and derived fields.
Geometries are stored as WKT/GeoJSON \textit{properties}.
Policies can consult these via predicates (e.g., \texttt{distance}<\(\tau\)) to express location-specific behavior.

\paragraph{Effects} Reaction rules may now attach external side-effects, such as launching an external service or opening a firewall rule.
Effects are tied to the rewrite itself---they occur when the graph is updated, not from guard evaluation---and are idempotent to support retries.

Together, properties and effects allow policies to integrate dynamic state and service instantiation directly.

\subsection{Spatial Names}\label{sec:spatial-names}

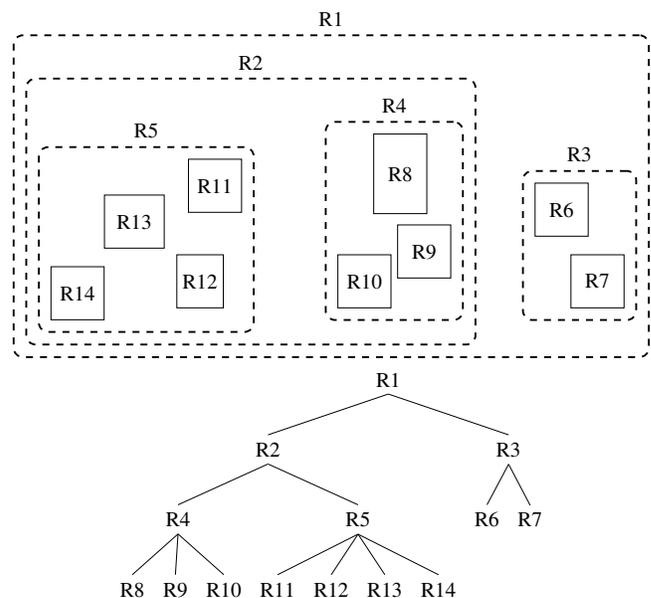
\begin{figure}[b!]
\centering
\begin{tikzpicture}[scale=0.9, every node/.style={font=\small},
  obj/.style={draw, rectangle, minimum width=0.8cm, minimum height=0.8cm, inner sep=2pt, align=center},
  mbrstyle/.style={draw, dashed, thick, rounded corners},
  treenode/.style={rectangle,draw,rounded corners,inner sep=6pt,minimum width=2.6cm,align=center},
  arr/.style={-{Latex[length=3mm,width=2mm]}}
]

\begin{scope}[local bounding box=plane]
\node[obj] (r14) at (1.2,1.0) {R14};
\node[obj, minimum width=0.9cm] (r13) at (2.15,2.2) {R13};
\node[obj, minimum width=0.7cm] (r12) at (3.25,1.2) {R12};
\node[obj, minimum width=0.8cm] (r11) at (3.5,2.8) {R11};

\node[obj] (r10) at (6.0,1.2) {R10};
\node[obj] (r9) at (7.0,1.7) {R9};
\node[obj, minimum height=1.2cm] (r8) at (6.6,3.0) {R8};

\node[obj] (r7) at (9.9,1.2) {R7};
\node[obj] (r6) at (9.3,2.4) {R6};

\node[mbrstyle, fit=(r14) (r13) (r12) (r11), inner sep=5pt, label={[name=r5l] above:R5}] (r5) {};
\node[mbrstyle, fit=(r10) (r9) (r8), inner sep=5pt, label={[name=r4l] above:R4}] (r4) {};
\node[mbrstyle, fit=(r7) (r6), inner sep=5pt, label={[name=r3l] above:R3}] (r3) {};

\node[mbrstyle, fit=(r5) (r5l) (r4) (r4l), inner sep=5pt, label={[name=r2l] above:R2}] (r2) {};
\node[mbrstyle, fit=(r3) (r3l) (r2) (r2l), inner sep=5pt, label={[name=r1l] above:R1}] (r1) {};
\end{scope}

\node[below=0 of plane] {
\Tree
[.R1
  [.R2
    [.R4 R8 R9 R10 ]
    [.R5 R11 R12 R13 R14 ]
  ]
  [.R3 R6 R7 ]
]
};

\end{tikzpicture}
\caption{R-tree spatial indexing enables efficient geometric queries.
  Spatial objects (R6--R14) are grouped hierarchically by minimum bounding rectangles, allowing logarithmic-time lookups.}
\label{fig:rtree}
\end{figure}

There is a broad class of network-connected device which derive their identity from their location.
The display in room 101 is fundamentally \textit{the room 101 display}---if the hardware breaks and gets replaced, the replacement assumes the same spatial identity; if the device moves to room 102, it becomes the room 102 display.
Spatial names~\cite{sns} combine functional hostnames (e.g., \texttt{display}, \texttt{mic}, \texttt{lock}) with a location hierarchy to create stable identities tied to physical spaces rather than to specific hardware.

Spatial names follow civic hierarchies that map naturally to DNS.
This hierarchy provides both human understanding and administrative delegation.
Names resolve differently based on context through split-horizon DNS: from within room 101, \texttt{display.room-101} resolves to local addresses such as link-local IP, Bluetooth, or Zigbee via extended resource records); from outside, it either resolves to a public IP address if explicitly exported or returns NXDOMAIN for privacy.
This enables local communication without WAN traversal while selective global exposure remains possible.

In our bigraph model, a device's position in the place graph directly determines its spatial name.
The containment relationship yields the DNS name through the place hierarchy.
When devices move and are reparented in the bigraph, their names update accordingly---no manual reconfiguration required.
This automatic binding means devices need only be physically placed to acquire their network identity.

To enable the ``point at devices'' interaction mentioned in \S\ref{sec:intro}, we augment bigraphs with spatial indexing.
Each node can carry geodetic metadata (coordinates, boundaries) as properties (\S\ref{sec:properties-effects}), and we maintain R-trees~\cite{guttmanRtreesDynamicIndex1984} over these geometries for efficient spatial queries.
As shown in Figure~\ref{fig:rtree}, R-trees hierarchically group nearby objects with minimum bounding rectangles, enabling logarithmic-time lookups for queries like ``which devices are in this view frustum?''.
When an augmented reality (AR) headset casts a ray or defines a viewing cone, the spatial index quickly returns candidate devices without traversing the entire graph.
This supports proximity-based discovery (``devices within 5 meters'') and geofencing (``entering room 101's boundary'').

\subsection{Spatially-Ephemeral Services}\label{sec:spatial-services}

Borrowing from just-in-time (JIT) instantiation~\cite{jitsu}, reaction rules can create \textit{spatially-scoped} microservices that exist only while certain conditions hold, and are dissolved when they no longer do.
These spatial microservices are \textit{ephemeral by construction}: their identity and lifetime are directly derived from the enclosing place.

We create spatial names for these services derived from the place hierarchy-- for example, \texttt{transcribe.room-101}---giving clients stable names even as backing instances are ephemeral.
The runtime can provision TLS certificates through a modified DNS server that exposes fine-grained capabilities via the Cap'n Proto capability-based remote procedure call (RPC) system~\cite{capnproto}.
When a spatial rule fires, it grants the ephemeral service a capability restricted to provisioning certificates for only its specific domain (e.g., \texttt{transcribe.room-101}).
This capability-based approach ensures services cannot obtain certificates for other domains, even if compromised.
The DNS server handles Automatic Certificate Management Environment (ACME) DNS-01 challenges internally and notifies the service via callbacks when certificates need renewal, triggering graceful restarts.
When the spatial guard retracts and the service terminates, it invokes the capability's teardown method to revoke the certificate and clean up DNS records.
When the meeting ends and the guard retracts, the name withdraws, the certificate expires, and the service vanishes---a clean lifecycle with no residual exposure or stale DNS entries.

A key consequence is that a service can be specified once at a global scope and realized as many per-place instances for a selected class of spaces (e.g., all meeting rooms on Floor~1), exported under a shared outer name to invited principals; each instance is scoped to its room and dissolves when the room becomes inactive.

This tight coupling of service lifecycle to spatial context yields the usual benefits of JIT instantiation---reduced idle cost and attack surface---alongside \textit{latency pre-purchase}: because instantiation is tied to spatial predicates (occupancy, schedule, proximity), effects can be scheduled ahead of time to mask expected setup delays.
For example, as soon as a user enters a building, their personal services (VPN, display layout, profiles) can be automatically pre-warmed near their likely destination.
Likewise, desks (or hotdesks) can configure themselves based on the identity and role of the occupant, with no manual binding to specific hardware.

Policies remain generally tied to roles and places, not device IDs, so new hardware inherits the same behavior automatically.
Spatial services are \textit{composable} across subgraphs and \textit{delegable} across boundaries via outer names, enabling rooms to cooperate within floors, and floors within buildings, without a central bottleneck.

\subsection{Implications of Spatial Policy}\label{sec:spatial-policy-implications}

We now detail the outcomes that arise from evaluating policy over space, and how they manifest in everyday applications as short, local policy.
Given rules attach to places and compose by nesting, without indirection, these policies are more concise and robust than those in existing automation stacks.
Rather than brittle choreographies of YAML automations, DNS/TLS plumbing, and other ad hoc integrations, authoring over \textit{space} shrinks configuration, increases reuse, and makes correctness easier to reason about.
Note that our model is not intended to replace existing automation platforms, but to provide a spatial substrate they can target.

\begin{itemize}
  \item\noindent{\textbf{Privacy by locality.}}
        Since bigraphs explicitly delimit which entities co-locate and which may interact, data and policies can be confined to local regions.
        Data only crosses boundaries through explicit links, so flows are intentional and governed by reaction rules.
  \item\noindent{\textbf{Robustness \& low-latency.}}
        Pattern-matching within a local subgraph ensures responsiveness.
        The model's robustness also benefits from its delegated and formally defined semantics.
        Subgraphs can operate autonomously without a single point of failure, and permit exhaustive reasoning about local behaviors.
  \item\noindent{\textbf{Correctness \& verification.}}
        Policies are expressed over a typed structure with clear invariants.
        This makes many safety properties expressible as spatial invariants, such as ``no untrusted user may enter PrivateSpace''.
        Formal verification techniques for bigraphs can be applied to check these invariants~\cite{formal_bigraphs}.
  \item\noindent{\textbf{Composability \& portability.}}
        A policy authored for one subgraph is parameterized only by its interface and labels, making it reusable across spaces.
        Composition scales naturally: nodes or regions can be added as sub-bigraphs, with existing policies merging automatically.
  \item\noindent{\textbf{Physical security.}}
        Requiring physical presence changes attack economics: virtual attacks scale globally at zero cost, but spatial attacks scale linearly with travel and risk.
        Attackers must be physically present, creating clear legal jurisdiction and forensic evidence.
\end{itemize}

The outcomes above manifest directly in practice.
Consider, for example, our private meeting scenario from \S\ref{sec:intro}.
Here, our model reduces the behavior to a single scoped rule: on entry, launch a transcription service on the local tablet, expose it under a place-derived name, and retract it on exit.
Privacy and latency come ``for free'' since evaluation is local and outer-name exposure is explicit; reuse holds because the rule travels to any space exporting the setup.
When the meeting ends, the service dissolves automatically, tying lifecycle to spatial guards.
As another example, \textit{hot-desking} similarly collapses to ``\texttt{User~$\in$~Desk}; apply profile; remove on leave'' eliminating ID binding so new hardware inherits policy simply by location without additional configuration.
These outcomes are hard to achieve with existing stacks: if-this-then-that (IFTTT) platforms
bind to node IDs (location is metadata), serverless/kubernetes react to API events behind global control planes (no spatial lifetime); and VLAN, Software-defined Networking, and Network Access Control encode reachability rather than \textit{where} policy holds.

\begin{figure*}[t!]
\begin{center}
\includegraphics[width=16cm]{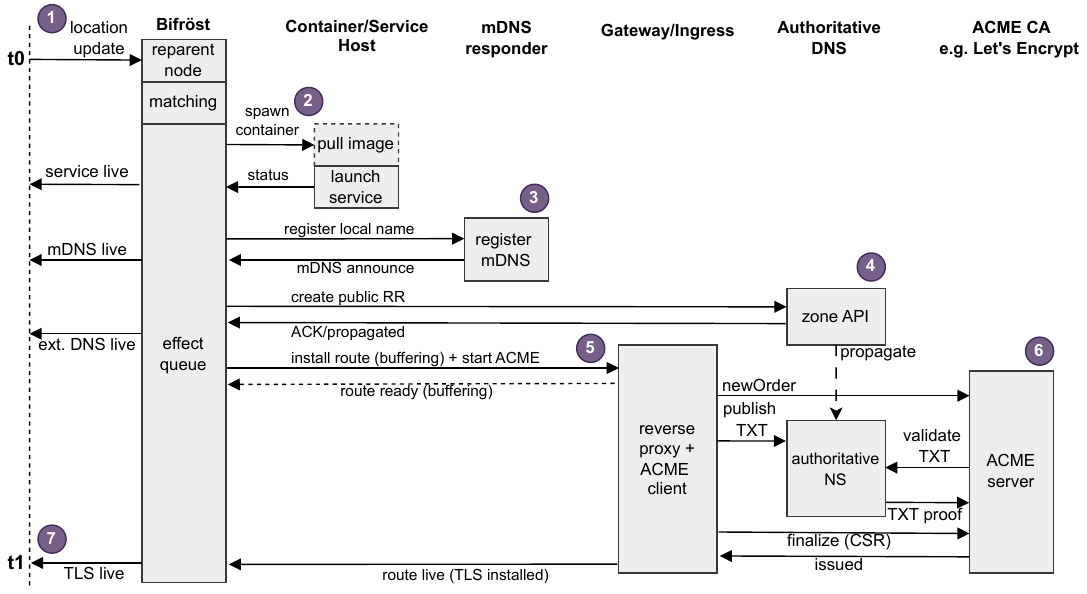}
\caption{Spatial guards mask time-to-ready latency.
  At $t_0$ the user enters a parent region (e.g., building).
  On location update (1), Bifr\"ost reparents the user, applies newly matching rules, and enqueues effects.
  These effects run while the user moves: the container/service is spawned on the host (2); a local name is advertised via mDNS (3); an external name is provisioned (4); and the gateway is configured to buffer traffic (5), then completes ACME validation and certificate issuance (6).
  By the time the user reaches the target space at $t_1$ (e.g., their office or a meeting room), the route is live with TLS and any early connects can be replayed, so the service is immediately available (7).
}\label{fig:swimlane}
\end{center}
\end{figure*}

\section{Spatial Networks}\label{sec:spatial-networks}

We now demonstrate our programming model through scenarios that expose fundamental tensions in today's systems: cloud platforms sacrifice locality for reach, while local hubs preserve autonomy but cannot compose across space.

\subsection{Digital Locks with Delegated Access}\label{sec:digital-locks}

Digital locks exemplify the need for local robustness with temporary delegated sharing.
Such locks should function offline, without reliance on a vendor cloud.
However, at the same time, users often require selective delegation---for example, an office may grant temporary access to a contractor or short-term guest.
Today, cloud-based locks enable such delegation, but at the expense of privacy and robustness: state changes are routed via vendor servers, usage data is exposed to providers, and outages can leave doors unusable.
Local hubs offer offline control, but enabling remote sharing requires complex setup with VPNs or exposed APIs, while revocation requires manual cleanup.
Scaling this to offices, campuses, or other large facilities typically means brittle VLAN segmentation.

Delegation with our model is simply a property update on the user node; a reaction scoped to the building attaches the guest's node to the lock while the temporal guard holds, automatically retracting the link when the window expires.
Presence-based unlocking is a containment or distance check in the graph, enforced locally.

Delegation and override become spatial guards, binding capabilities to place and time and scaling from homes to campuses.
When network boundaries map to physical boundaries, attackers must be physically present to compromise systems; physical containment makes network isolation intrinsic to the environment.

\subsection{Bridging Local and Global Zones}

Many deployments require the ability to effectively bridge environments with global reach, where presence in one zone grants access to secure, direct communication or control channels in another, without relying on fragile device firmware.
Today, cloud-based platforms offer global APIs but not spatial scoping: services are bound to user or device accounts, not to places, and every action incurs a round-trip through a remote control plane.
Local hubs can provide zone isolation using VLANs or subnets, but bridging zones demands custom VPNs alongside scripts that poll location changes and reconfigure tunnels---effortful, error-prone, and non-composable across deployments.

With our model, the link graph encodes potential peering.
When a user enters zone~A and is reparented under its subgraph, a reaction establishes the relevant peering to zone~B, opening access only to services permitted by spatial policy (e.g., a display or chat channel).
When the user exits, the aperture dissolves with the rule, leaving no residual exposure.
The large IPv6 address space allows each zone its own prefix; then, over a lightweight VPN tunnel, a user can---during a meeting with colleagues tunneled from another office---share content from their laptop to a display located in the remote office’s conference room.
Authentication can rely on physical presence, ensuring that only participants physically in the office can share or view content, and only while the meeting lasts.
This can be approximated with existing tools using static VPNs and manual configuration, but such solutions are brittle, labor-intensive, and lack portability.

\subsection{Audio Challenge-Response \\ for Co-Location}

Physical boundaries as network boundaries enable security protocols that rely on physical co-location.
The audio challenge-response protocol~\cite{2005_ieee_audio} exemplifies this: a device plays an ultrasonic tone, and responding devices must prove they heard it by incorporating it into their cryptographic response.
By adding this as a rule guard, this establishes that responders are present in the same room; devices in other rooms cannot participate as they cannot hear the challenge.

Without room-level network boundaries, this protocol fails.
On a building-wide network, devices in another room could potentially participate in a room authentication despite being physically separated.
Cloud platforms make this worse: the challenge must traverse WANs, exposing it to replay attacks and making physical proximity meaningless.
With spatial networking, the challenge-response happens within the room's network boundary.
The physical containment of the room naturally enforces the security property the protocol requires.

\subsection{Audio Transcription in Meetings}

The meeting scenario from \S\ref{sec:intro} requires transcription that is fast, private, and ephemerally tied to occupancy (starting only when quorum is reached, blanking immediately on intrusion, and extending seamlessly to remote participants when rooms are linked).
Today, cloud-based transcription provides ease of use but routes audio through external servers, introducing privacy risks, WAN latency, and Internet dependence.
Local hubs can run transcription engines on-premises, but extending them across sites (e.g., distant offices) requires manually exposing services and managing per-participant encryption keys.

Our model can instead instantiate transcription as a spatially ephemeral service under the meeting-room node, triggered by a guard over authorized participants.
Participant keys are drawn from properties on their nodes, and audio is encrypted to the entire set before processing.
If zones are linked, the service is exported under a temporary outer name, enabling remote participation without broad exposure.
Intrusions are detected locally, with sub-millisecond reaction times (as shown in \S\ref{sec:impl}); when the meeting ends, the service name, aperture, and keys dissolve automatically.
In effect, privacy and robustness emerge directly from spatial scope, while global sharing remains possible under explicit links.
This level of coordination---quorum-sensitive, encrypted, multi-room transcription---is beyond today's clouds without compromising privacy, and beyond today's hubs without fragile, hand-crafted, and non-portable integrations.

\section{Implementation}\label{sec:impl}

We implemented Bifr\"ost as a practical runtime for spatial networking, departing from verification-oriented bigraph tools to prioritize real-time coordination.
Our OCaml library uses mutable data structures for sub-millisecond reactions (\S\ref{sec:impl-search}), exports bigraphs via Cap'n Proto RPC for delegated operation across hosts (\S\ref{sec:delegated-runtime}), and provides Python bindings for integration with existing automation stacks.

\subsection{Library \& Matching}
\label{sec:impl-search}

Bifr\"ost is an OCaml library with a runtime for delegated operation and language interoperability.
The core provides a mutable bigraph with typed nodes, properties, and effects; rules match spatial patterns and apply rewrites whose side-effects enact configuration changes (\S\ref{sec:properties-effects}).

Each node has a unique integer identifier bound to a record carrying its control (with compile-time type), a vector of ports, a \textit{mutable parent reference} (for the place graph), a dictionary of properties, and optional human-readable labels.
The place graph is a forest realized via parent references, supporting constant-time containment checks and re-parenting; children are stored as linked sets for efficient iteration.
The link graph is a hypergraph realized as a bidirectional map: ports reference link identifiers, while link identifiers index member ports.
This enables fast membership queries and pattern matching without global traversal, and supports efficient serialization into Cap'n Proto~\cite{capnproto} messages.
Properties are a typed key–value dictionary; guards operate on these (e.g., sensor readings, authorization flags).
We use OCaml polymorphic variants to enforce type checks dynamically while keeping rule specifications concise.

Reactions are applied by in-place updates of parent \textit{references} and link memberships; effects are typed commands attached to rewrites, executed atomically, idempotent for retries, and with rollback to ensure resilience.
The runtime locates redex \textit{embeddings}---structure-preserving mappings of the redex's nodes and ports into the host graph---and then applies the corresponding reaction.
We support two complementary modes.
\textit{Targeted matching} addresses a specific node or set of nodes by stable keys (e.g., a node ID, label, or outer name).
The matcher anchors on such keys, verifies the surrounding place/link/interface constraints, and halts after a complete set of matches.
This avoids global search and minimizes latency for real-time reactions.
By contrast, \textit{portable matching} applies rules wherever a structural pattern holds (e.g., ``all rooms with an HVAC unit''), matching by controls/types/properties rather than IDs.
Backtracking depth-first search extends a partial mapping while first checking place constraints (parent/child, root/site interfaces) and only then link constraints (port incidence).
Candidate host nodes are prefiltered by control and arity; the enumerator is \textit{streaming}, yielding embeddings lazily without materializing the full set in memory.
These indices and heuristics (type/arity prefilters, place-first checks, port-degree heuristics) prune aggressively; worst-case complexity remains exponential in the redex size.
However, applying a match touches only the matched region, keeping application time effectively flat as delegated deployments grow.

\subsection{Delegated Runtime \&  \\ Language Interoperability}\label{sec:delegated-runtime}

The core library is usable in delegated environments, and includes a runtime that directly exports graphs and rules via Cap'n Proto RPC.
This runtime has a dual role: (1) it provides Cap'n Proto interfaces for reading, querying, and updating remote subgraphs, and (2) it evaluates rules and dispatches their external effects.


We implement a flat Cap'n Proto schema; each bigraph is serialized as: (i) a dense array of node records with control, ports, label, and property map; (ii) a parent array encoding the place forest; and (iii) link memberships mapping hyperedges to their member ports and ports back to hyperedges.
This adjacency-list encoding avoids deep nesting, ensures serialization cost grows linearly with $|V| \Plus |E|$, and permits zero-copy deserialisation.
We use Cap'n Proto over alternatives such as JSON or Protobuf precisely for its compactness, zero-copy streaming, and capability-oriented RPC.

The runtime also supports registering callbacks via Cap'n Proto RPC, providing APIs for subscription and mutation.
Synchronization across delegated hosts is scoped explicitly by outer names: when a node exports an open link, the runtime publishes updates to the peer process holding the adjoining subgraph.
This design allows subgraphs to operate autonomously while still being composable into larger ones.
This enables offloading computation; individual devices can manage and evolve their local subgraphs; full state is propagated upstream only when necessary.
Spatial composition boundaries are explicit and policy-driven: developers can specify which links and nodes are visible, and govern aggregation.
Because synchronization is initiated by open links, nodes can operate offline; the local graph remains authoritative for its interior.

The runtime includes Python bindings in order to support integration with its wider ecosystem of machine learning and automation frameworks.
The bindings wrap Cap'n Proto messages in idiomatic Python objects.
Python clients can inspect subgraphs, subscribe to updates, and inject new rules back into the runtime.
Developers write simple Python callbacks; the runtime ensures delegated consistency and executes effects, while external automation logic (e.g., policy engines, LLMs) runs outside the core.

\section{Evaluation}\label{sec:eval}

Our evaluation aims to assess whether (1) local reactions remain fast and stable as deployments grow, (2) delegated operation introduces acceptable overheads and (3) end-to-end service behavior is practical.
We assess both the latency and memory overheads of local rule applications for increasing graph size and varying pattern shape.
For delegated synchronization, the overheads added by our model are the (de)serialization and merging of subgraph updates.
We also evaluate an end-to-end delegated application across a real networked office graph, and record the breakdown from trigger to service-ready.

\noindent{\textbf{Setup:}} Microbenchmarks ran on a 64-bit GNU/Linux host (x86\_64, 8 cores, 16 GB RAM; native release build).
The macrobenchmark ran the controller/runtime on the same host, while container start-up and effects executed on a RPi 3 Model B (Cortex-A53, 1 GB RAM, 100 Mb/s Ethernet), emulating a low-power hub.

\begin{figure}[!htbp]
  \centering
  \includegraphics[width=8.5cm]{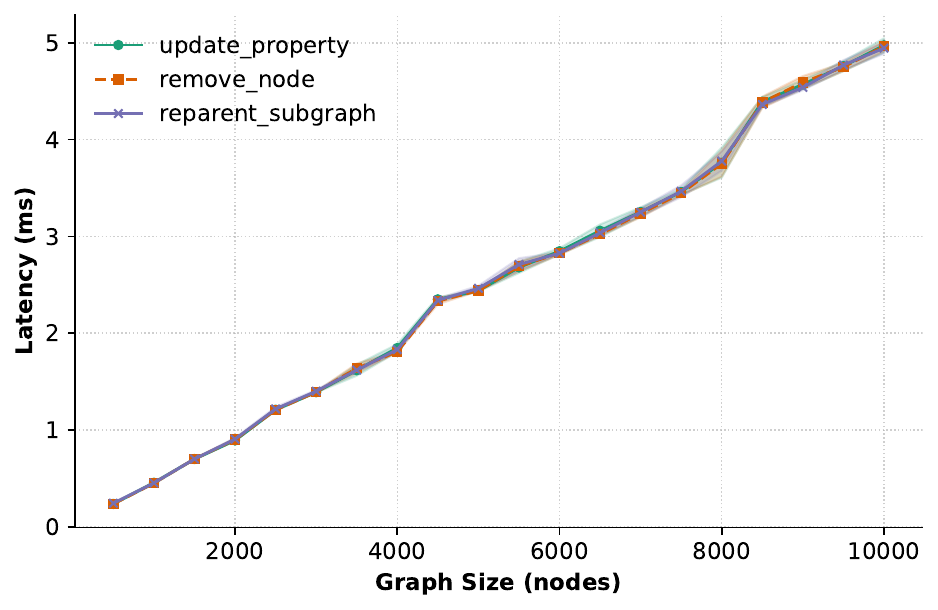}
  \caption{Plot of time taken to apply reactions \textit{vs.} the number of graph nodes.
    Measured over 100 runs for each graph with 95\% CIs shown.}
  \label{fig:rule_app_latency}
\end{figure}

\subsection{Rule Application Overheads}

We measure latency as the sum of search and application time, scaling only the background graph size; the matched pattern remains fixed in size and degree.
We stress the runtime by streaming and enumerating \textit{all} candidate embeddings before applying the first match (i.e. \textit{portable matching} from \S\ref{sec:impl-search}).
Workloads cover representative updates exercised by our runtime: updating a node property; removing a node from the place graph; and reparenting a subgraph, as used when we merge subgraphs in delegated operation.
The background graphs grow by composing additional subgraphs so degree delegated and control/label frequencies stay stable as $|V|$ increases.
Match patterns are placed uniformly at random to avoid placement bias.
We run 100 trials per workload and size, report medians with 95\% confidence intervals, and exclude load/deserialize from the latency metric.

Figure~\ref{fig:rule_app_latency} shows end-to-end latency grows \textit{approximately linearly} with $|V|$.
Growth is dominated by search: even with indices over controls and labels that aggressively prune candidates (\S\ref{sec:impl}), enumeration still 
touches more regions with growing deployments.
The actual application latency remains effectively flat across growing graphs since updates are localized (pointer/adjacency edits on the matched region only).
In absolute terms, latency stays in the millisecond range, with application time sub-millisecond even for large (10\textsuperscript{4} node) backgrounds.

The result is conservative, since enumeration is a worst case used for global matches.
Note that 10\textsuperscript{4} nodes reflects a campus-scale upper bound; in practice, deployments should be delegated so per-place latency is far lower.

Peak memory use during search and application tops out at <50MB; with a lazy enumerator, usage is  bounded by the footprint required to hold the currently loaded subgraph.
This is suitable for constrained hardware.

\subsection{Operational Overheads}

\begin{figure}[t]
  \centering
  \includegraphics[width=8.5cm]{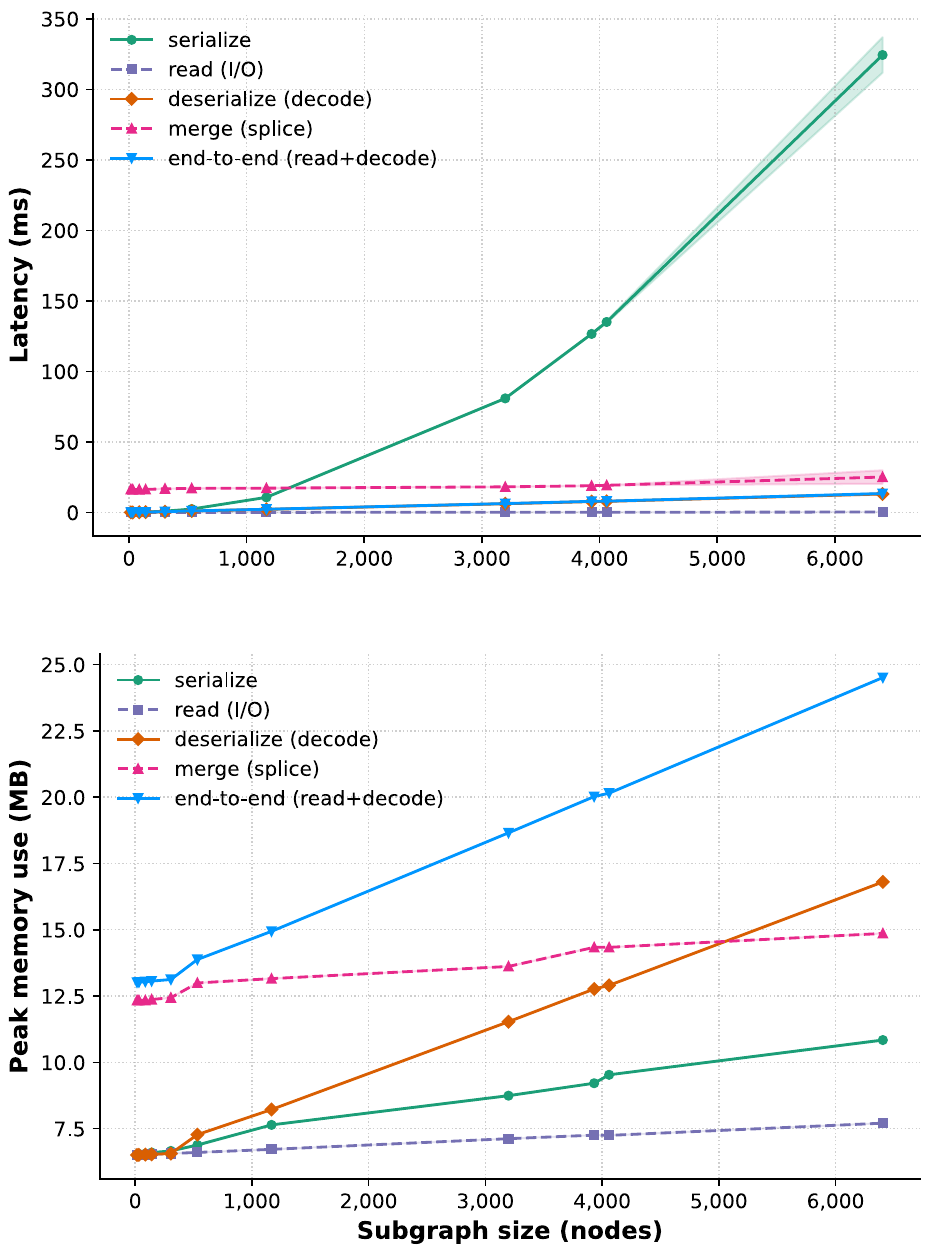}
  \caption{Plot of time taken and memory required to serialize, deserialize, read, and merge a subgraph \textit{vs.} the number of subgraph nodes.
    Measured over 100 runs for each subgraph with 95\% CIs shown.}
  \label{fig:serdes_latency}
\end{figure}

We evaluate the cost of delegated subgraphs across hosts by measuring latency/memory across four stages, end-to-end and in isolation: serialization (encode to bytes), read, deserialization (decode), and merge (splice the decoded subgraph into a host graph).
These stages mirror delegated operation: shipping a bounded place subgraph across an outer-name boundary and replacing the corresponding subtree on the receiver.
We vary only the \textit{subgraph} size being shipped; the surrounding ``campus'' graph is held fixed.
For each target size $k$, we select the smallest subtree with $\ge k$ nodes and run 100 trials.
We report medians with 95\% confidence intervals.
Peak memory (Figure~\ref{fig:serdes_latency}, top) grows roughly linearly and remains small (<50 MB): deserialization memory use grows steadily, merge remains relatively flat, and serialization and read are modest.
Latency is ms-scale (Figure~\ref{fig:rule_app_latency}, bottom): \textit{serialization dominates} overall cost and grows with subgraph size.
In contrast, read, deserialization, and merge remain essentially flat; we use a precomputed parent–child index, with only a small dependence on root fan-out.

\subsection{Spatially-Scoped Meeting Transcription}\label{sec:impl-demo}

\begin{figure}[htbp]
  \centering
  \includegraphics[width=5.5cm]{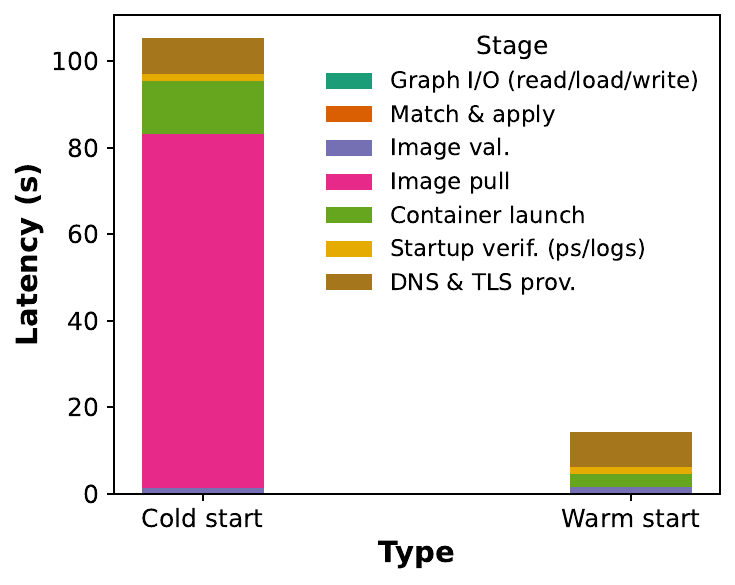}
  \caption{End-to-end latency from reaction to service availability, broken down by graph processing, matching \& application, image handling, container startup, and post-start verification.}
  \label{fig:e2e_latency}
\end{figure}

We built out the meeting transcription use case as an end-to-end application in our office, atop a real model of our space.
On entry, a user's GPS updates from OwnTracks are reverse-geocoded to an OSM \texttt{way\_id}; a reaction attached to the corresponding place reparents the user node and fires.
The effect spawns a Docker container on a separate host running an audio transcription model optimized for real-time streaming, configures split-horizon DNS (a local mDNS name plus a short-TTL external alias), and provisions short-lived TLS certificates.
This validates end-to-end integration on a real office graph: place resolution and guard matching, container launch, naming and credential issuance, and aperture management across linked rooms.
In our measurements (Figure ~\ref{fig:e2e_latency}), reaction time remains sub-millisecond; end-to-end readiness is dominated by container setup, and the overheads from graph processing, including reparenting, are negligible.

\section{Related Work}\label{sec:related}


\paragraph{Early ``smart-space'' operating systems}
Early smart-space systems framed a physical room as a first-class computing entity.
\textit{Gaia} cast an ``active space'' as the unit of abstraction, with OS-like services for discovery, context, and application composition so applications could treat an environment as a programmable substrate~\cite{gaia}.
\textit{iROS}, developed for interactive workspaces, similarly tied the abstraction to a room but emphasized decoupled coordination via the ``event heap'', a tuplespace-style service that tolerated device churn and ad hoc collaboration~\cite{iros,eventheap}.
Both demonstrated that treating a \textit{space}---not just a host---as the locus of computation simplified multi-device experiences.
Bifr\"ost generalizes this idea with a formal substrate that composes across nested regions and derives routing, naming, and capability scopes directly from spatial structure, providing guarantees that earlier room-centric OSes left implicit.

Later works extended these ideas to home environments.
\textit{HomeOS} unified heterogeneous devices with role-based APIs for portable ``home apps''~\cite{homeos}, while \textit{SafeHome} added transactional semantics for automations on a local hub~\cite{safehome}.
These platforms demonstrate the robustness of local hubs but retain a centralized, home-wide model.
Bifr\"ost instead links naming, policy, and service instantiation with spatial relations, yielding capability-scoped services.

\paragraph{Delegated coordination and programming models} Related work explored coordination models for dynamic environments.
\textit{One.world} offered tuples, events, discovery, and migration so applications could adapt as devices and users moved~\cite{oneworld}.
Borcea et al. use smart messages to let developers target computations to physical spaces~\cite{spatialprog}.
These models highlight proximity and mobility, but treat space largely as metadata rather than a first-class substrate.
Bifr\"ost re-centers space in the programming model, explicitly encoding containment and connectivity.

\paragraph{IoT automation}
\textit{ParaDrop} demonstrated that gateways and access points could host containerized services close to devices, reducing latency and cloud dependence~\cite{paradrop}.
Early attempts to further localize data processing pursued a similar vision: user-owned data containers and building-level platforms that emphasized privacy and autonomy~\cite{databox, hat}.
However, these efforts predated today's rich ecosystem of device APIs and integrations.
At the time, few consumer devices exposed programmable interfaces, making it difficult to achieve compelling applications beyond demos.
In contrast, modern hubs (e.g., Home Assistant, openHAB) can federate thousands of APIs across commercial ecosystems.
However, their IFTTT-style rules operate over node IDs rather than space.
Emerging standards, like Matter~\cite{matter} over Thread~\cite{thread}, improve onboarding and interoperability but still treat location as metadata.
This creates the critical mass of programmability that makes spatially scoped networking not only viable but necessary: the bottleneck is no longer device interoperability, but the lack of abstractions to capture physical boundaries as first-class networking constructs.

\paragraph{Ephemeral services, naming, and spatial models}
Just-in-time instantiation has been explored in various works, from Jitsu's unikernel instantiation to open serverless platforms~\cite{jitsu,openlambda}.
Bifr\"ost couples instantiation to \textit{spatial relations}, alongside network events.
DNS-SD and mDNS standardize discovery but lack explicit spatial or temporal scope~\cite{rfc6763,rfc6762}; hierarchical location names with split-horizon resolution reconcile local and global views~\cite{sns}.
Our ephemeral, place-derived names are compatible with these systems while enforcing bounded exposure.

\paragraph{Bigraph tooling and verification}
Bigraphical reactive systems, and related calculi such as mobile ambients, foreground where computation happens~\cite{milner2008bigraphs,cardelli}.
BigraphER is a mature OCaml library that 
targets exhaustive state-space exploration and solver-backed verification~\cite{bigrapher}.
Our aim is complementary: a runtime substrate for spatial coordination.
We implement record mutability, extend the formalism with typed properties and idempotent effects, and expose delegated operation via Cap'n Proto RPC.
These choices trade exhaustive verification for 
practical coordination.

\paragraph{Spatially-aware computing in fiction}
Science fiction has long envisioned computation that disappears into the physical environment.
Vinge's \textit{Rainbow's End} depicts computing woven into clothing and viewed through contact lenses, creating information-rich augmented reality tied to physical locations.
Stross's \textit{Halting State} imagines ambient computation where every mobile phone contributes processing power to ``the Zone''---a distributed environment where augmented reality and physical space merge, with game logic executing on the devices of whoever happens to be nearby.
\textit{Star Trek}'s starships exemplify ambient intelligence: crew members speak to the air and the computer responds based on their location and identity.
Doors recognize who may pass, replicators remember preferences, and the computer routes communications based on spatial context.
Iain M. Banks' \textit{Culture} series pushes further: ship ``Minds'' maintain omnipresent computational awareness, growing furniture from floors when needed, adjusting gravity locally, and creating privacy fields.
Every surface can become responsive, every space computationally active, with citizens connected through neural laces to this pervasive intelligence.
These works of fiction paint a vision where computation becomes a property of space itself, responding to who is present, understanding what they need, and adapting based on where they are.
Bifr\"ost provides the networking nuts and bolts to make parts of this fiction reality.

\section{Conclusions}\label{sec:conclusions}

We detail \textit{Bifr\"ost}, a spatial coordination runtime grounded in bigraphs.
Our runtime augments bigraphs with typed properties and idempotent effects, enabling lightweight handlers to react to spatial events and enact safe, effectful updates.
The runtime is natively delegated; it instantiates spatially ephemeral services under place-derived names and executes rewrites with co-located effects.
This reframes networking itself: physical boundaries become network boundaries, ephemeral services materialize as routable network entities, and compose across rooms or buildings compiles down to explicit overlay links.
We detail how such spatial scoping yields low-latency, robust, and private coordination bridging local and global scopes.

Various directions follow naturally from this work:

\paragraph{Integration with existing automation stacks}
Rather than usurp today's automation platforms, \textit{Bifr\"ost} acts as an effect backend (i.e. spatial policies could compile to Home Assistant or openHAB integrations, while retaining place-derived ephemeral names for cross-boundary exposure).
Building out such integration preserves device diversity and driver support, while ensuring that spatial scoping can be incrementally deployed in heterogeneous environments.

\paragraph{Coordination with embodied agents}
LLM-based agents can operate autonomously over \textit{Bifr\"ost}, effectively authoring reactions using a deliberately small, capability-limited toolset.
The result is anticipatory yet safe behavior (e.g., a meeting concierge that pre-warms services from calendar context) while keeping observation and actuation fully confined to place and outer-name boundaries, in line with least-privilege exposure.
Such deployment is now practical: low-power neural accelerators provide the headroom for efficient reasoning~\cite{bulp}, while advances in agent capabilities---long-horizon spatial planning~\cite{planning} and tool use~\cite{tool_use}---enable operation in complex, dynamic spaces.

\paragraph{Safety and verification for agent-authored policy}
Tooling and verification should carefully bound an agent's scope and actions, with actions validated against spatial invariants before any effects fire, and logs for effects emitted across boundaries.
We envision runtime guards for agent-synthesized reactions: policies should be statically checked for capability typing, outer-name exposure bounds, and adherence to runtime locality constraints.
This yields a balance between autonomy and verifiability.

\paragraph{Large-scale environmental monitoring} Spatial reasoning also extends to devices that are not co-located with dense infrastructure, such as environmental monitoring deployments in remote or transboundary regions.
Here, connectivity may be intermittent, yet coordination across organizational boundaries still requires fine-grained spatial scoping.
\textit{Bifr\"ost} enables such devices to securely sign and scope their data at the point of capture, linking measurements to spatial metadata and capability rather than raw identifiers.
Real-time processing can then remain local---on the device or a nearby gateway---preserving privacy and autonomy without mandatory cloud dependence, while delay-tolerant dissemination ensures eventual consistency across larger regions.
This approach provides both resilience and trust: organizations can interoperate without replicating raw repositories or hardcoding APIs, and each reading carries attested provenance tied to its spatial context.

\paragraph{Spatial service deployment}
Spatial services require deployment mechanisms that go beyond traditional containerization.
While Docker enables our prototype, it lacks principled approaches to secrets management, dynamic reconfiguration, and spatial auto-scaling---services should spawn additional instances when rooms fill or migrate compute as users move between spaces.
We envision extending our capability-based model beyond DNS and certificates to encompass the full service lifecycle: capabilities for storage provisioning, secret rotation, and cross-machine coordination.
This would enable self-managing spatial systems where services not only spawn based on spatial predicates but also heal from failures, scale with occupancy, and migrate seamlessly as the physical environment evolves.

\newpage

\bibliographystyle{plain}
\bibliography{reference}

\end{document}